\documentclass[12pt,3p]{elsarticle}

\usepackage{amsmath,amsfonts,amsthm,amssymb,indentfirst,epic,url,graphicx,needspace}

\usepackage{moreverb,url}
\usepackage{subfigure, float}
\usepackage{multicol}
\usepackage{threeparttable}

\usepackage[colorlinks,bookmarksopen,bookmarksnumbered,citecolor=red,urlcolor=red]{hyperref}

\newcommand{\bbeta}{ \mbox{\boldmath $ \beta $} }

\newcommand{\balpha}{ \mbox{\boldmath $ \alpha $} }

\newcommand{\bsigma}{ \mbox{\boldmath $\sigma$} }

\newcommand{\bSigma}{ \mbox{\boldmath $\Sigma$} }

\newcommand{\bLambda}{ \mbox{\boldmath $\Lambda$} }

\newcommand{\bgamma}{ \mbox{\boldmath $\gamma$} }
\newcommand{\brho}{ \mbox{\boldmath $\rho$} }

\newcommand{\bdelta}{ \mbox{\boldmath $\delta$} }
\newcommand{\bOmega}{ \mbox{\boldmath $\Omega$} }

\newcommand{\N}{\mathcal{N}}

\newcommand{\bzero}{\textbf{0}}

\newcommand{\bI}{\textbf{I}}

\newcommand{\bt}{\textbf{t}}

\newcommand{\bx}{\textbf{x}}

\newcommand{\by}{\textbf{y}}

\newcommand{\bz}{\textbf{z}}

\begin{document}
\begin{frontmatter}


\title{Bayesian latent time joint mixed effect models for multicohort longitudinal data}
\author[label1]{Dan Li}
\address[label1]{Alzheimer's Therapeutic Research Institute, University of Southern California, San Diego, USA}

\author[label1,label2]{Samuel Iddi}
\address[label2]{Department of Statistics, University of Ghana, Legon-Accra, Ghana}

\author[label3]{Wesley K. Thompson}
\address[label3]{Department of Psychiatry, University of California, San Diego, USA}

\author[label1]{Michael C. Donohue}
\ead{mdonohue@usc.edu}

\author{for the Alzheimer's Disease Neuroimaging Initiative\fnref{labelADNI}}
\fntext[labelADNI]{Data used in preparation of this article were obtained from the Alzheimer's Disease Neuroimaging Initiative (ADNI) database (adni.loni.usc.edu). As such, the investigators within the ADNI contributed to the design and implementation of ADNI and/or provided data but did not participate in analysis or writing of this report. A complete listing of ADNI investigators can be found at: \url{http://adni.loni.usc.edu/wp-content/uploads/how_to_apply/ADNI_Acknowledgement_List.pdf}}

\begin{abstract}
Characterization of long-term disease dynamics, from disease-free to end-stage, is integral to understanding the course of neurodegenerative diseases such as Parkinson's and Alzheimer's; and ultimately, how best to intervene. Natural history studies typically recruit multiple cohorts at different stages of disease and follow them longitudinally for a relatively short period of time. We propose a latent time joint mixed effects model to characterize long-term disease dynamics using this short-term data. Markov chain Monte Carlo methods are proposed for estimation, model selection, and inference. We apply the model to detailed simulation studies and data from the Alzheimer's Disease Neuroimaging Initiative.
\end{abstract}


\end{frontmatter}

\section{Introduction}
\label{sec1}
Disentangling the effects of normal aging from the effects of pathophysiology on neurobiological trajectories is crucial for predicting who will develop Alzheimer's or other neurodegenerative diseases. The determination of neurobiological trajectories is typically based on data obtained over a relatively short time frame relative to the time span of normal aging and neurodegeneration \cite{Thompson(2011)}. There are four key barriers to the accurate characterization of longitudinal trajectories of neurodegeneration and pathology in older adults: 1) late-life trajectories can span 40 or more years; 2) there is substantial variation in the timing and temporal dynamics of late-life trajectories; 3) study selection criteria and sampling have a substantially different impact across age groups; and 4) there are potentially significant levels of censoring due to death and disability in late-life that cannot be treated as independent of pathology trajectories in most cases. Accounting for these issues is crucial to studying the onset, course and fluctuations of neurodegenerative diseases \cite{Jack(2010), Jack(2013)}, and for guiding when and how to target interventions and preventative approaches \cite{Sperling(2012)}. A better understanding of how and when to intervene necessitates the determination of dynamic networks of trajectories of markers of cognition, function, and pathology. Long-term disease dynamics are of great interest and importance, and have been hypothesized without rigorous methods. 

Methods for longitudinal data analysis are ubiquitous. For example, generalized linear or nonlinear mixed-effects models {\cite{Laird_Ware(1982), Lindstrom(1990)}} are commonly applied to model growth curves of height or weight over time since a meaningful ``time zero'' (e.g. birth). However, neurodegenerative diseases like Alzheimer's and Parkinson's that progress over long periods of time are typically studied by taking longitudinal samples of multiple cohorts at different stages of disease {\cite{Petersen(2010), Marek(2011)}}. The multiple cohorts have no common meaningful time zero. Time since dementia diagnosis might be a candidate time zero, however the transition to dementia is censored for many individuals in the pre-dementia cohorts. Also, the dementia diagnosis is subjective and may vary from one clinician to the next in a multicenter study. Further, dementia is not an absorbing state and reversion to pre-dementia diagnoses occur.

We propose a latent time joint mixed effects model (LTJMM) for characterizing biomarker trajectories in aging. The model extends joint mixed effects models \cite{Laird_Ware(1982)} to include an individual-specific latent time shift. Each individual's latent time shift is shared across all of their outcomes and represents the extent of their long-term disease progression. The model is similar to others proposed for analysis of the Alzheimer's Disease Neuroimaging Initiative (ADNI)\cite{Petersen(2010)} (e.g. Jedynak et al.\cite{Jedynak(2012)} and Donohue et al.\cite{Donohue(2014)}). However, the proposed model can also accommodate covariates for fixed-effects and the Bayesian implementation allows flexible but rigorous interrogation of the posterior distribution to make inferences about long-term disease dynamics and the potential propagation of treatment effects from early biomarkers to downstream cognitive and functional measures. Estimation and inference are accomplished using Hamiltonian Monte Carlo sampling as implemented in \textsf{Stan} \cite{stan(2016)} and the \textsf{R} package \textsf{rstan} (Stan Development Team\cite{rstan(2016)}, version 2.10.1). The \textsf{Stan} code model specifications and related tools for estimation and prediction are available as an additional \textsf{R} package from \url{https://bitbucket.org/mdonohue/ltjmm}.

\section{Methods}
\label{sec2}

\subsection{Latent time mixed-effects models for joint longitudinal responses}
\label{model}

Suppose that $p$ response variables are measured for $n$ individuals at different follow-up times. Follow-up may differ from subject to subject and outcome to outcome, so we denote the measured outcome $k$ for individual $i$ at time $j$ as $y_{ijk}$, where $i=1,\ldots,n$, $k=1,...,p$ and $j=1,\ldots,q_{ik}$. The $p$ measured outcomes could be mixtures of binary, count, ordinal and continuous. Assuming $y_{ijk}$ has a distribution in the exponential family with canonical parameter $\theta_{ijk}$, the latent time joint mixed effects model (LTJMM) is given by
\begin{equation}
	\label{eq:1}
	h_k\left(\theta_{ijk}\right) = \eta_{ijk} = \bx^\prime_{t_{ijk}} \bbeta_k+ \gamma_k \left(t_{ijk} + \delta_i\right) + \alpha_{0ik} + \alpha_{1ik} t_{ijk} + \varepsilon_{ijk},
\end{equation}
where $h_k\left(\cdot\right)$ is a monotonic differentiable link function specific to the $k$-th outcome, $\eta_{ijk}$ is the linear predictor for the $k$-th outcome from individual $i$ at time $j$.  The vector $\bx^\prime_{t_{ijk}}$ represents possibly time-varying covariates, and the vector $\bbeta_k$ is their corresponding regression coefficients.  The $\varepsilon_{ijk} \sim \N\left(0, \sigma_k^2 \right)$ is a measurement error term, which accounts for outcome-specific variance.  

The parameters $\alpha_{0ik}$ and $\alpha_{1ik}$ are the subject and outcome specific random intercept and slope, and $\delta_i$ is the subject-specific time shift shared among outcomes. ``Short-term'' observation time is represented by observed covariate $t_{ijk}$. The parameter $\gamma_k$ corresponds to the outcome-specific slope with respect to ``shifted'' or ``long-term'' time $t_{ijk}+\delta_i$. The time shift $\delta_i$ quantifies the progression of the $i$-th individual relative to the population, which is assumed to follow $\delta_i\sim \N\left(0, \sigma_{\delta}^2\right)$. And $\alpha_{1ik}$ provides the information on whether individual $i$ is evolving faster or slower than the average individual for outcome $k$.

When applying the model, we typically include a fixed effect for age, as a time-varying covariate, for each outcome. This admits two relevant rate-of-change parameters: one for biological age and one for ``disease time''. This reflects the fact that individuals can reach the same stage of disease at different ages, and allows independent effects of healthy aging and disease progression. The random effects accommodate additional subject to subject variation in the level and rate of change of disease markers. The link function, $h_k$, can be used to accommodate any distribution from the exponential family.

A constraint must be placed on (\ref{eq:1}) to ensure model identifiability based on the observed data without relying on informative prior distributions. In particular, to ensure identifiability of $\bdelta$, the following constraint on random intercepts is sufficient: $\sum_{k=1}^{p}\alpha_{0ik}=0$ for all individuals $i=1,\ldots,n$. A proof of identifiability is available in the supplemental material. Model (\ref{eq:1}) can be modified to address more flexible nonlinear relationships over time by incorporating higher-order terms or splines \cite{Ruppert(2003)}, provided these extensions maintain monotonicity. Shape invariant models and self modeling regression (e.g. Kneip and Gasser\cite{Alois(1998)}) similarly require positive first derivatives. Under model (\ref{eq:1}), we trivially assume that at least one of the outcome-specific slopes $\gamma_k$ is not equal to zero, $k=1,\ldots,p$. In our application, all the outcomes are oriented to be increasing, and thus we assume all $\gamma_k>0$, $k=1,\ldots,p$.  

Finally, we explore two different distribution assumptions for random effects, namely:
\begin{description}
	\item[] 1. Univariate Gaussian random effects:
	$\alpha_{0ik}\sim \N\left(0,\sigma_{0k}^2\right)$ and 
	$\alpha_{1ik}\sim \N\left(0,\sigma_{1k}^2\right)$;
	\item[] 2. Multivariate Gaussian random effects:
	$\balpha_i \sim \N\left(\bzero, \bSigma \right)$,
\end{description}
where $\bSigma$ is a covariance matrix of dimension $2p$. The latter assumption allows more direct exploration and inference regarding the correlation of response variables.

\subsection{Prior specification}

Since an improper prior may result in an improper posterior, we will use proper but weakly informative priors on all the model parameters \cite{Gelman(2006)}. The regression parameters $\bbeta$ and $\gamma$ are assigned independent weakly informative normal $\N\left(0, 100\right)$ priors (truncated below by 0 for $\bgamma$).  Without specific information, the half-Cauchy prior is a good default choice for scale parameters \cite{Gelman(2006)}. The standard deviations, i.e., $\sigma_\delta$, $\sigma_{01}, \cdots, \sigma_{0p}$, $\sigma_{11}, \cdots, \sigma_{1p}$ and $\bsigma$, are given weakly informative half-Cauchy priors with a small scale, i.e., $\text{half-Cauchy}\left(0, 2.5\right)$.  The prior variance 100 for the regression parameters is chosen to be sufficiently high to be vague enough, but sufficiently low to avoid slow mixing due to near impropriety of the posterior \cite{Natarajan(1998)}.

To ensure efficiency and arithmetic stability, we applied the Cholesky decomposition to the covariance matrix for the random effects, $\bSigma$, allowing us to model the standard deviations and correlations independently.  We first decomposed the covariance matrix as $\bSigma=\bLambda \bOmega \bLambda$, where $\bLambda$ is a diagonal matrix with diagonal elements $\sigma_{01},\cdots,\sigma_{0p}$, $\sigma_{11},\cdots,\sigma_{1p}$, and $\bOmega$ is the correlation matrix with 1's on the diagonal and off-diagonal elements $\brho$.  A Choleksy decomposition gives $\bOmega=\boldsymbol{L}\boldsymbol{L}^{\prime}$, where $\boldsymbol{L}$ is a lower triangular matrix. Furthermore, $\bSigma=\bLambda\boldsymbol{L}\boldsymbol{L}^{\prime}\bLambda$. During sampling, a draw is obtained from a multivariate Gaussian $\bz \sim \N\left(\bzero,\bI\right)$, and then the random effects are calculated as $\bLambda \boldsymbol{L} \bz$. Here, $\bz$ is regarded as a random reparameterization and independent of $\brho$. As suggested in \textsf{Stan} \cite{stanmanual(2016)} manual, we imposed an LKJ prior on the correlation matrix \cite{Lewandowski(2009)}.

\subsection{Model comparison criteria}

We use two model comparison criteria, namely, the widely applicable information criterion (WAIC) \cite{Watanabe(2010)} and the leave-one-out cross-validation information criterion (LOOIC) \cite{Gelfand(1992), Gelfand(1996)}.  WAIC and LOOIC can be computed using the log-likelihood evaluated at the posterior simulations of the parameter values.  Both have various advantages over the deviance information criterion (DIC).  WAIC can be viewed as an improvement on DIC for Bayesian methods \cite{Vehtari(2016)}. We refer to Gelman et al.\cite{Gelman(2014)} for the detailed rationale for the preference of WAIC to DIC. As WAIC is reported on the deviance scale \cite{Gelman(2014)}, a difference in WAIC value of $2-6$ is considered as positive evidence, $6-10$ strong evidence, and $>10$ very strong evidence \cite{Kass(1995)}. 

Approximate LOO can be computed using raw importance sampling (IS, Gelfand et al.\cite{Gelfand(1992)}).  Vehtari et al.\cite{Vehtari(2016)} improved the LOO estimate using Pareto smoothed importance sampling (PSIS) method which provides a more accurate and reliable estimate than IS. The \textsf{R} package \textsf{loo} \cite{loopackage} provides tools for efficient computation of WAIC and LOOIC. The minimum WAIC and LOOIC indicate the best fit.

\section{Results}
\label{sec3}

\subsection{Simulation study}
\label{simulation_study}

We conducted a simulation study to assess the model performance and assess small sample bias.  Data were generated from two candidate LTJMMs with identity link: (M1) Univariate Gaussian random effects; (M2) Multivariate Gaussian random effects. We simulated $n=100$ individuals with $p=4$ outcomes and $q=4$ time points each.  The observation times for each individual were sampled from a Uniform distribution $t_{ijk} \sim \text{Uniform}\left(0, 10\right)$.  Additional simulation parameters were set to $\bbeta=\left(1, 0.5, 2, 0.8\right)^{\prime}$, $\bgamma=\left(0.2, 0.1, 0.25, 0.5\right)^{\prime}$, $\sigma_\delta=4$, $\bsigma=\left(0.1, 0.2, 0.3, 0.25\right)^{\prime}$,  $\bsigma_{\alpha_0}=\left(0.5, 1, 0.8\right)^{\prime}$, and $\bsigma_{\alpha_1}=\left(1, 2, 1.5, 1\right)^{\prime}$.

Let $\hat{\theta}_m$ denote the posterior estimate of a parameter $\theta$ for the $m$-th simulation and let $M$ denote the total number of simulations. We consider the following quantities to assess model performance: (i) total bias, $\text{Bias}=\sum_{m=1}^{M}\left(\hat{\theta}_m-\theta_m\right)/M$, (ii) total mean squared error of prediction, $\text{MSPE}=\sum_{m=1}^{M}\left(\hat{\theta}_m-\theta_m\right)^2/M$, and (iii) coverage rate of the 95\% credible intervals, $C_{95}$. Results are reported and discussed in Section \ref{simulation_results}. Each model was simulated
$M=100$ times and both models were fit to each simulated data set.

For each model fit, two parallel Markov chains were run with dispersed initial values to diagnosis convergence.  For each single chain, we ran $2\:000$ iterations and discarded the first $1\:000$ iterations as a warm-up phase, yielding a total of $2\:000$ samples for posterior analysis.  The remaining samples were used to calculate posterior summaries of the parameters of interest and model comparison measurements.  The potential scale reduction statistic $\hat{R}$ \cite{Gelman(1992)}, calculated by \textsf{Stan} was used to verify posterior convergence.

\subsection{Simulation study results}
\label{simulation_results}

\begin{figure}[H]
	\centering
	\subfigure[]{\includegraphics[scale=0.6]{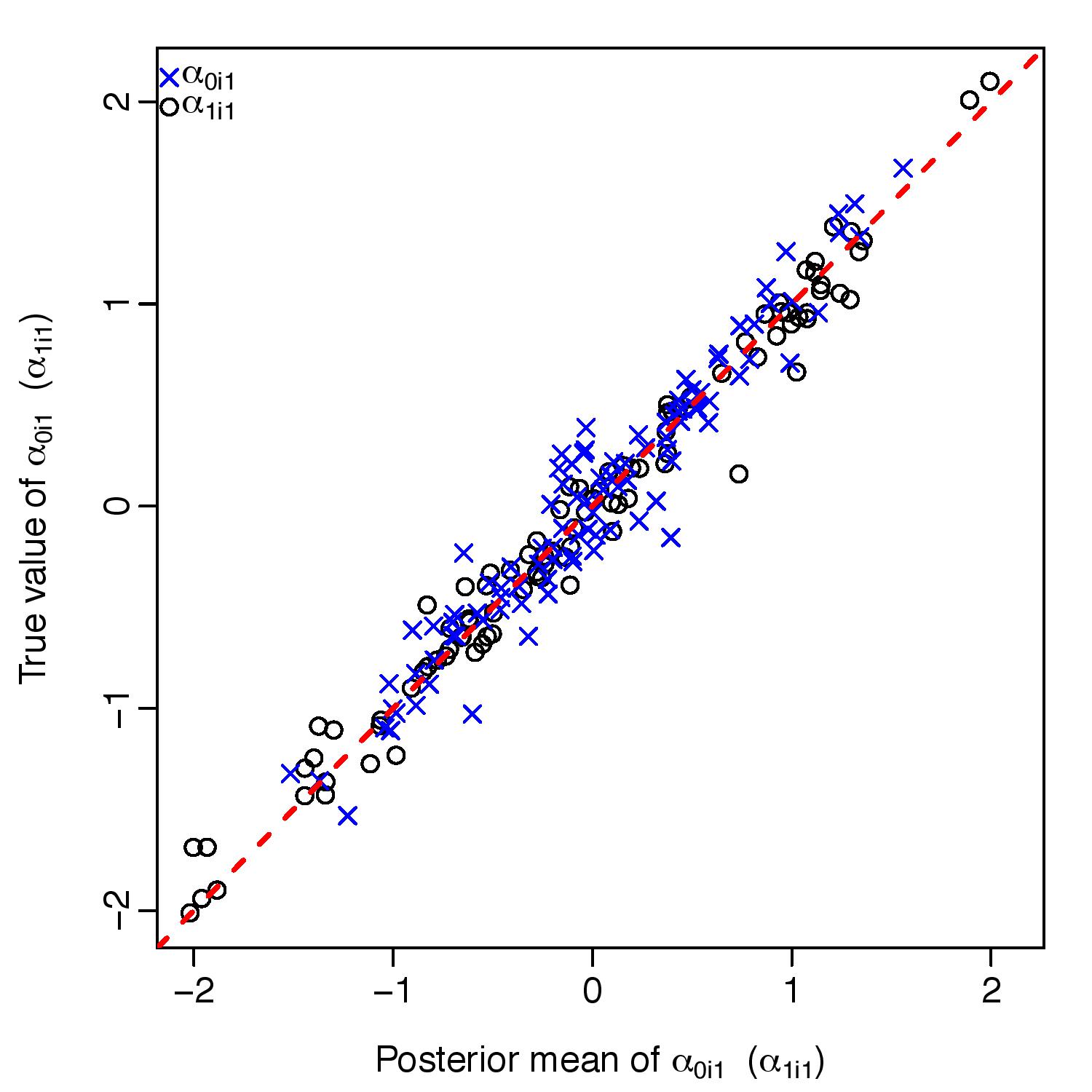}}
	\subfigure[]{\includegraphics[scale=0.6]{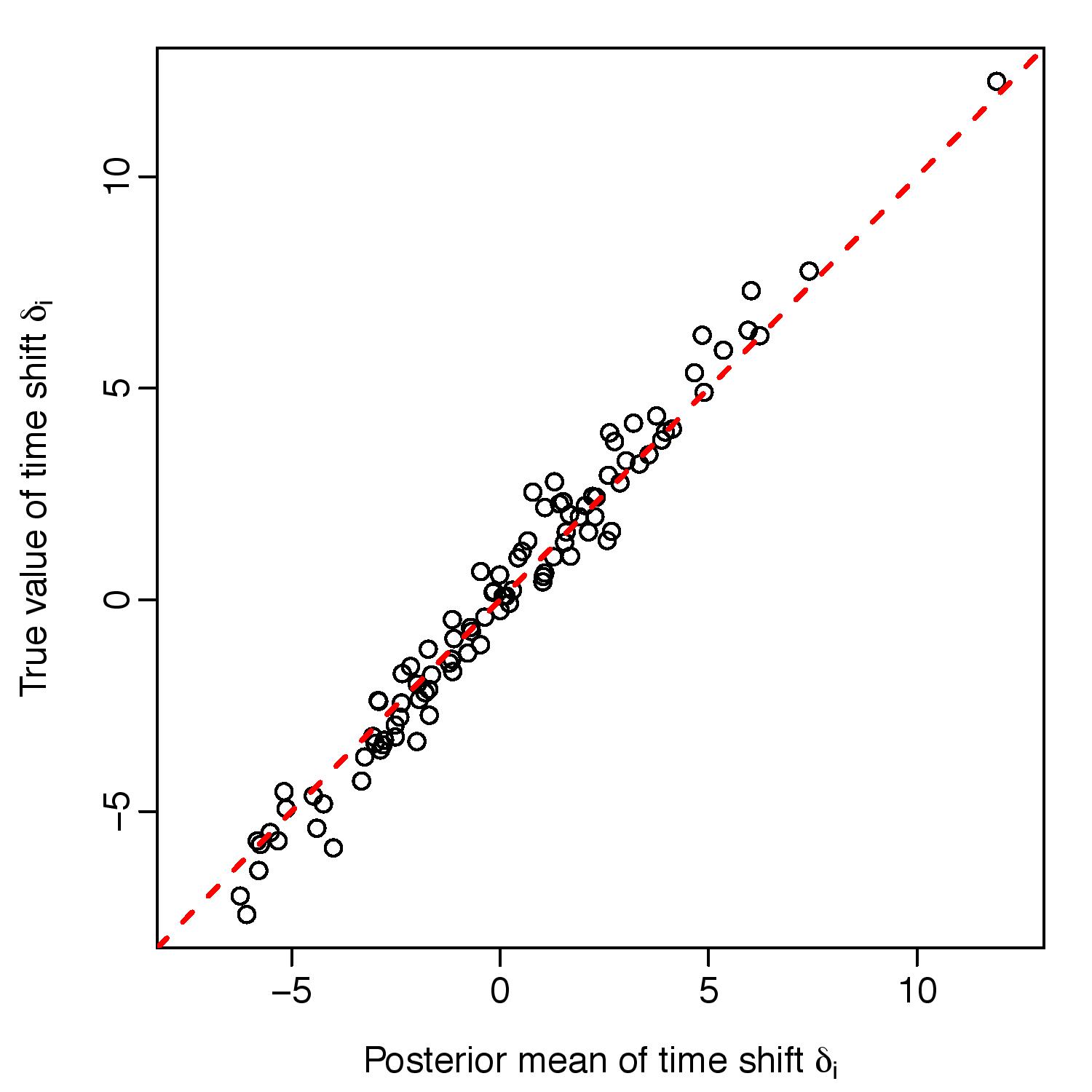}} \\ \vspace{-5pt}
	\caption{Results of one simulated data example with 4 outcomes, 100 individuals, and at 4 time points. (a) True value versus posterior mean of random intercepts and slopes for the first outcome. Black circle and blue cross indicate the slope and intercept, respectively. (b) True value and posterior mean of latent time shifts, and these are indicated in black circles. The red dashed line is an identity line (points lie close to the line indicating a good fit). \label{Fig1}}
\end{figure}

The estimated potential scale reduction factors were below 1.1 for all parameters, indicating successful convergence. Table \ref{table:results_simulation} summarizes the measures of total bias, MSPE, $C_{95}$, and the model comparison criteria for each scenario. When the data are generated from M2, the true model is most often preferred in terms of either LOOIC (96\%) or WAIC (81\%). The quartiles of the comparison criteria differences between M2 and M1 are significantly negative, indicating strong evidence that M2 outperforms M1. When the data are generated from M1, the models are similar, and the comparison criteria are performing as expected and guiding us to the true model on average (LOOIC: 59\% and WAIC: 65\%). Models with criteria differences smaller than two might be considered equivalent. This is expected since M2 accommodates the association between outcomes and includes M1 as a special case. For both scenarios, most parameters of the true model are estimated with better $C_{95}$, lower bias and MSPE, indicating good performance in terms of prediction error.  Figure \ref{Fig1} plots the true versus estimated individual intercepts and slopes for the first outcome, and the true versus estimated time shifts of one simulated data set. Both plots demonstrate agreement between the true and estimated values.

\begin{table}[H]
	\small\sf\centering
	\renewcommand{\arraystretch}{1}
	\caption{Simulation study results. One hundred simulated data sets were generated under model M1 (univariate Gaussian random effects) and model M2 (multivariate Gaussian random effects). Both models were fit to all 200 data sets. \label{table:results_simulation}}
	\vspace{8pt}
	{\tabcolsep=1pt
		\begin{tabular}{@{}c|ccc|ccc|ccc|ccc@{}}
			\hline
			& \multicolumn{6}{c|}{True Model: M1} & \multicolumn{6}{c}{True Model: M2} \\ & \multicolumn{3}{c|}{Fitted Model: M1} & \multicolumn{3}{c|}{Fitted Model: M2} & \multicolumn{3}{c|}{Fitted Model: M1} & \multicolumn{3}{c}{Fitted Model: M2} \\
			Parameter & Bias & MSPE & $C_{95}$ & Bias & MSPE & $C_{95}$ & Bias & MSPE & $C_{95}$ & Bias & MSPE & $C_{95}$\\ \hline
			$\beta_1$ & -0.0022 & 0.0086 & 0.96 & -0.0005 & 0.0087 & 0.96 & -0.0013 & 0.0064 & 0.96 & -0.0016 & 0.0062 & 0.98\\
			$\beta_2$ & -0.0043 & 0.0131 & 0.93 & 0.0003 & 0.0131 & 0.95 & 0.0006 & 0.0133 & 0.94 & 0.0020 & 0.0131 & 0.94\\
			$\beta_3$ & -0.0059 & 0.0157 & 0.97 & -0.0061 & 0.0160 & 0.97 & -0.0041 & 0.0155 & 0.97 & -0.0044 & 0.0147 & 0.97\\
			$\beta_4$ & -0.0425 & 0.0481 & 0.97 & -0.0409 & 0.0477 & 0.97 & -0.0428 & 0.0550 & 0.95 & -0.0420 & 0.0556 & 0.97\\
			$\gamma_1$ & 0.0274 & 0.0025 & 0.93 & 0.0321 & 0.0027 & 0.90 & 0.0305 & 0.0036 & 0.83 & 0.0449 & 0.0044 & 0.89\\
			$\gamma_2$ & 0.0194 & 0.0016 & 0.96 & 0.0222 & 0.0018 & 0.95 & 0.0246 & 0.0025 & 0.93 & 0.0360 & 0.0033 & 0.93\\
			$\gamma_3$ & 0.0265 & 0.0031 & 0.96 & 0.0324 & 0.0033 & 0.97 & 0.0388 & 0.0061 & 0.83 & 0.0551 & 0.0067 & 0.92\\
			$\gamma_4$ & 0.0398 & 0.0082 & 0.96 & 0.0504 & 0.0087 & 0.95 & 0.0529 & 0.0134 & 0.85 & 0.0621 & 0.0102 & 0.95\\
			$\sigma_1$ & 0.0006 & $<.0001$ & 0.91 & 0.0006 & $<.0001$ & 0.92 & 0.0006 & $<.0001$ & 0.92 & 0.0007 & $<.0001$ & 0.91\\
			$\sigma_2$ & 0.0020 & $<.0001$ & 0.95 & 0.0021 & $<.0001$ & 0.94 & 0.0009 & $<.0001$ & 0.97 & 0.0020 & $<.0001$ & 0.94\\
			$\sigma_3$ & 0.0023 & 0.0002 & 0.94 & 0.0028 & 0.0002 & 0.93 & -0.0001 & 0.0002 & 0.93 & 0.0021 & 0.0002 & 0.94\\
			$\sigma_4$ & 0.0018 & 0.0002 & 0.96 & 0.0022 & 0.0002 & 0.96 & 0.0028 & 0.0002 & 0.96 & 0.0021 & 0.0002 & 0.96\\
			$\sigma_{\delta}$ & -0.2383 & 0.4542 & 0.93 & -0.3312 & 0.4846 & 0.91 & -0.2502 & 0.7656 & 0.84 & -0.4459 & 0.6715 & 0.89\\
			$\sigma_{01}$ & 0.0142 & 0.0037 & 0.93 & 0.0336 & 0.0048 & 0.87 & 0.0162 & 0.0021 & 0.91 & 0.0185 & 0.0022 & 0.89\\
			$\sigma_{02}$ & 0.0061 & 0.0052 & 0.97 & 0.0313 & 0.0064 & 0.95 & 0.0298 & 0.0057 & 0.96 & 0.0258 & 0.0055 & 0.95\\
			$\sigma_{03}$ & 0.0033 & 0.0060 & 0.94 & 0.0255 & 0.0067 & 0.94 & 0.0262 & 0.0058 & 0.91 & 0.0147 & 0.0051 & 0.96\\
			$\sigma_{11}$ & 0.0078 & 0.0066 & 0.94 & 0.0333 & 0.0079 & 0.91 & 0.0067 & 0.0066 & 0.89 & 0.0093 & 0.0061 & 0.93\\
			$\sigma_{12}$ & 0.0114 & 0.0095 & 0.97 & 0.0462 & 0.0119 & 0.94 & 0.0122 & 0.0206 & 0.94 & -0.0089 & 0.0183 & 0.95\\
			$\sigma_{13}$ & 0.0003 & 0.0088 & 0.95 & 0.0297 & 0.0101 & 0.92 & -0.0014 & 0.0111 & 0.97 & -0.0283 & 0.0104 & 0.98\\
			$\sigma_{14}$ & 0.0105 & 0.0053 & 0.96 & 0.0356 & 0.0067 & 0.93 & 0.0114 & 0.0064 & 0.97 & 0.0076 & 0.0056 & 0.97\\
			\% best LOOIC & \multicolumn{3}{c|}{59\%} & \multicolumn{3}{c|}{41\%} & \multicolumn{3}{c|}{4\%} & \multicolumn{3}{c}{96\%}\\
			\% best WAIC & \multicolumn{3}{c|}{65\%} & \multicolumn{3}{c|}{35\%} & \multicolumn{3}{c|}{19\%} & \multicolumn{3}{c}{81\%}\\ \hline
			\multicolumn{13}{c}{Quartiles $\mathrm{\left(Q_1,Q_2,Q_3\right)}$ of the criteria differences between M2 and M1 }\\ \hline
			LOOIC $\mathrm{diff_{M2-M1}}$ & \multicolumn{6}{c|}{$\left(-4.7313, 2.2414, 9.6980\right)$} & \multicolumn{6}{c}{$\left(-36.2904,-24.3240,-16.4950 \right)$}\\
			WAIC $\mathrm{diff_{M2-M1}}$ & \multicolumn{6}{c|}{$\left(-2.0153, 2.5876, 7.4679 \right)$} & \multicolumn{6}{c}{$\left(-12.0306, -6.6466, -1.2454 \right)$}\\ \hline
		\end{tabular}
		\begin{tablenotes}
			\footnotesize
			\item Note: The best performance was determined by the lowest WAIC and LOOIC among the two models for each data and summarized as ``\% best'' over the 100 simulations. The lower quartile, median and upper quartile are respectively denoted by $\mathrm{Q_1}$, $\mathrm{Q2}$ and $\mathrm{Q3}$.
		\end{tablenotes}
	}
	\end{table}

\subsection{Application to Alzheimer's Disease Neuroimaging Initiative}
\label{ADNI_application}

The LTJMM was fit to data from the Alzheimer's Disease Neuroimaging Initiative (ADNI), which has followed volunteers diagnosed as cognitively healthy or with varying degrees of cognitive impairment since 2005 \cite{Petersen(2010)}. The ADNI battery includes serial neuroimaging, cerebrospinal fluid (CSF), and other biomarkers; and clinical and neuropsychological assessments. Participants returned for repeated assessments at six months, one year, and every year thereafter. Seven outcome measures were included in the model: CSF tau and amyloid beta (A$\beta$) 1-42; PET imaging of amyloid deposition and glucose metabolism in the brain; volumetric magnetic resonance imaging (vMRI) of the hippocampus; the 13 item Alzheimer's Disease Assessment Scale (ADAS13); and the Functional Activities Questionnaire (FAQ). Fixed effect covariates for each outcome included age, carriage of the APOE$\varepsilon$4 allele, sex, and education. The model was fit with and without assuming a multivariate Gaussian distribution for the random effects. Three parallel Markov chains were run for $8\:000$ iterations and the first $4\:000$ warm-up iterations were discarded. Every fourth value of the remaining part of each chain was stored to reduce correlation, yielding a total of $3\:000$ samples for posterior analysis.  The posterior mean and the 95\% credible intervals were calculated using the obtained samples for each parameter.

One goal of the analysis is to compare the long-term trends of the outcomes on a comparable scale and make conclusions about the temporal ordering of their emergence. With this goal in mind, the outcome measures were transformed first to quantiles, and the quantiles were then transformed by the inverse Gaussian quantile function. All the transformed outcomes were then modeled as Gaussian with identity link. ADNI subjects were diagnosed at their first visit with normal cognition (24\%), subjective memory concern (6\%), early mild cognitive impairment (17\%), late mild cognitive impairment (32\%), and mild to moderate dementia (19\%). A weighted quantile transformation \cite{Hmisc} was used so that quantiles approximate the quantiles from a sample with equal numbers of each diagnosis. Categorical diagnosis is based, in part, on the subjective interpretation of the clinical presentation (excluding CSF and imaging data) by the physician, and hence was not included as a covariate in LTJMMs.

We fit the LTJMMs to examine gains by incorporating correlated random effects. Furthermore, we compare our proposed LTJMMs with existing models: the commonly used linear mixed effects model (MM) with and without fixed effects for categorical baseline diagnosis and its interaction with time. Table {\ref{results_adni_b}} summarizes the model comparison results. We found that $\textrm{LTJMM}_2$ was preferred over $\textrm{LTJMM}_1$ and MM without diagnosis. However, MM with diagnosis outperformed our proposed $\textrm{LTJMM}_2$ in terms of WAIC, LOOIC and DIC.

\begin{figure}[H]
	\centering
	\includegraphics[scale=1]{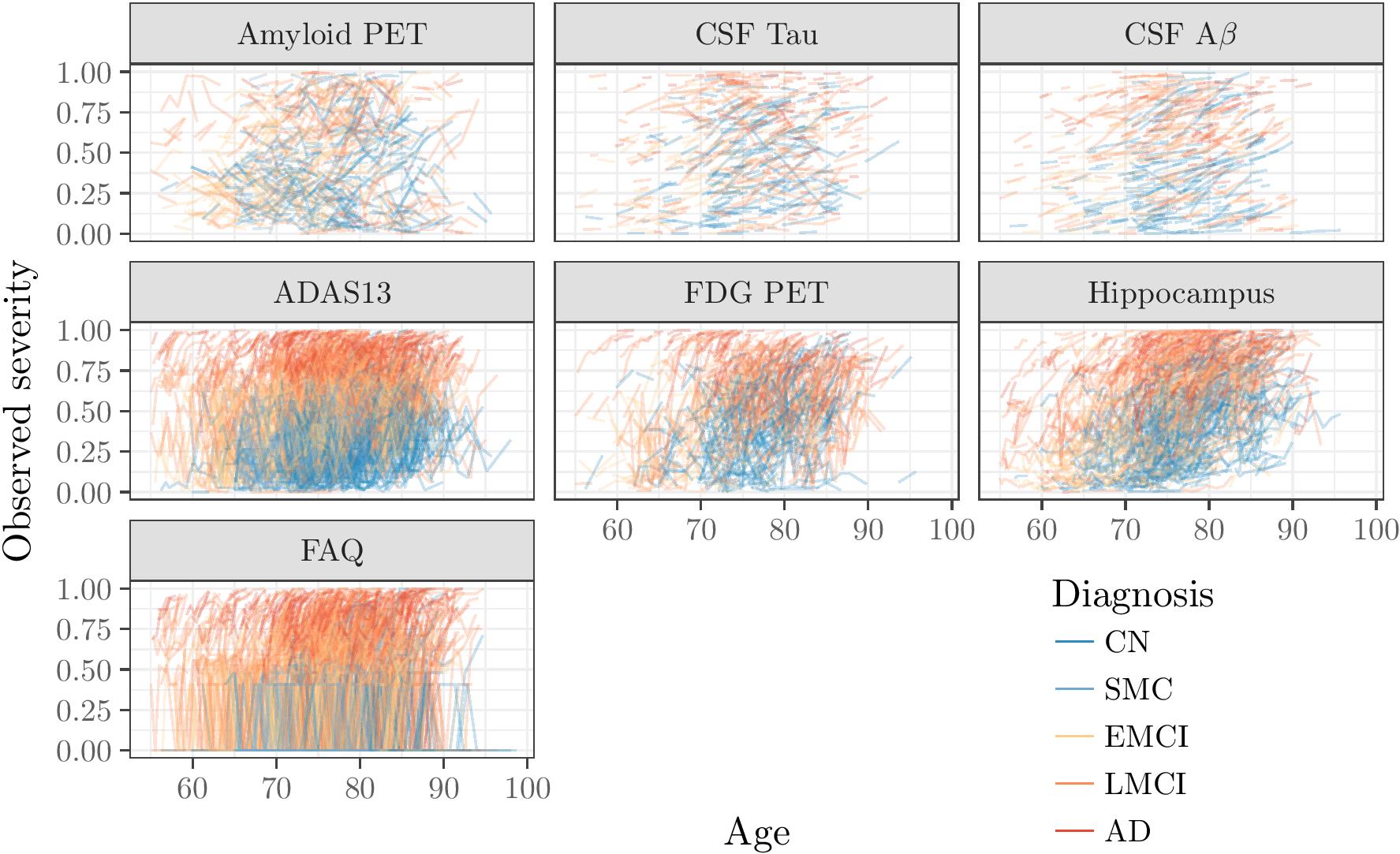}\\
	\vspace{10pt}
	\includegraphics[scale=1]{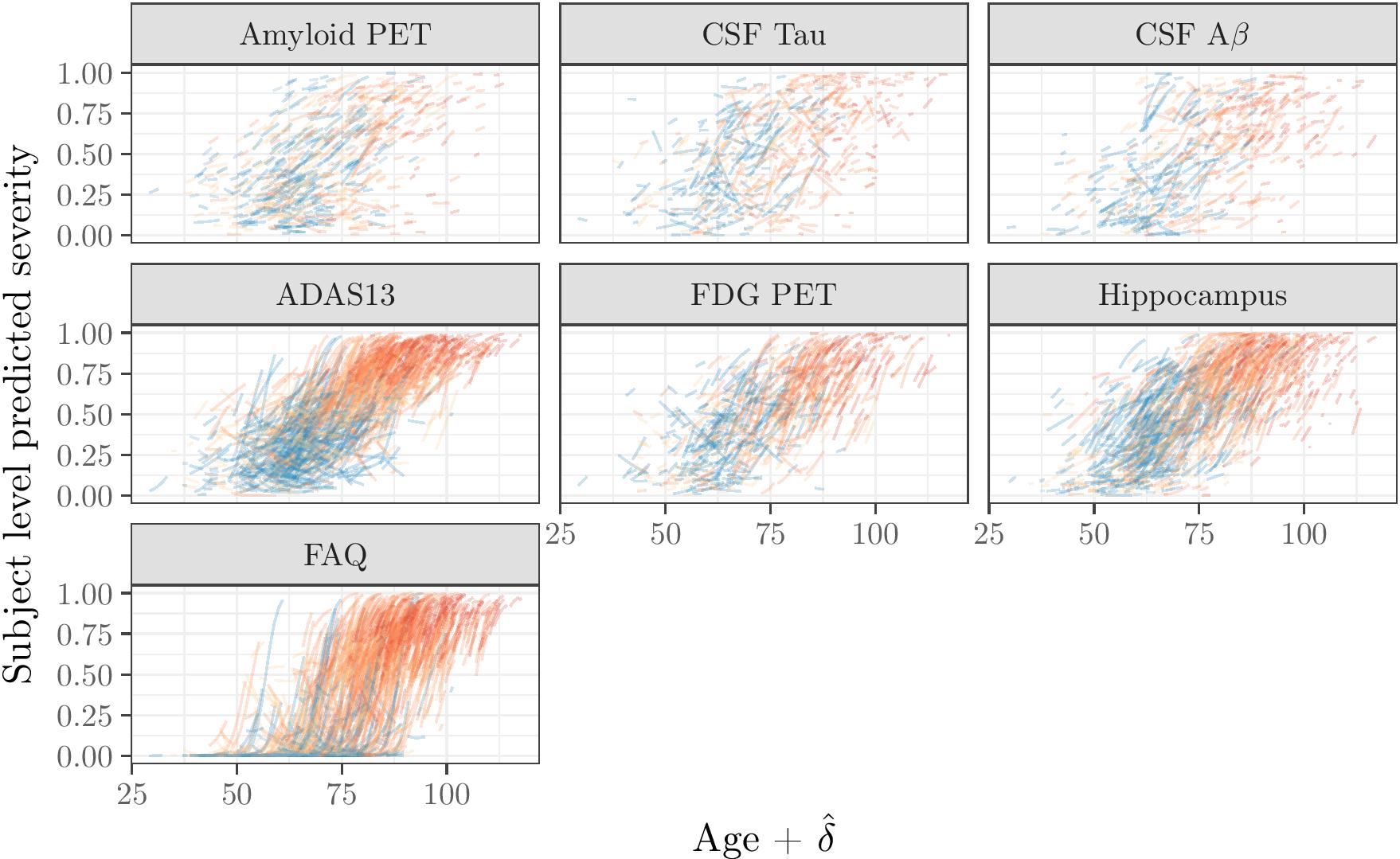}
	\caption[]{Subject-level observed and predicted severity. The top panel shows spaghetti plots of the observed quantiles of each outcome from all subjects in the Alzheimer's Disease Neuroimaging Initiative with respect to their age over time. The bottom panels show modeled trajectories for these same subjects from the fitted LTJMM with respect to the sum of age and estimated latent time, $\delta$. The colors indicate diagnostic severity at first observation, from cognitively normal (blue) through dementia (red).\\
		\footnotesize{Abbreviations: ADAS13, Alzheimer's Disease Assessment Scale (13 Item version); FDG, fluorodeoxyglucose; PET, positron emission tomography; CSF, cerebrospinal fluid; FAQ, Functional Activities Questionnaire; CN, cognitively normal; SMC, subjective memory concern; EMCI, early mild cognitive impairment; LMCI, late mild cognitive impairment; AD, probable Alzheimer's Disease with mild to moderate dementia.} \label{spaghetti}}
\end{figure}

\begin{figure}[H]
	\centering
	\includegraphics[scale=0.8]{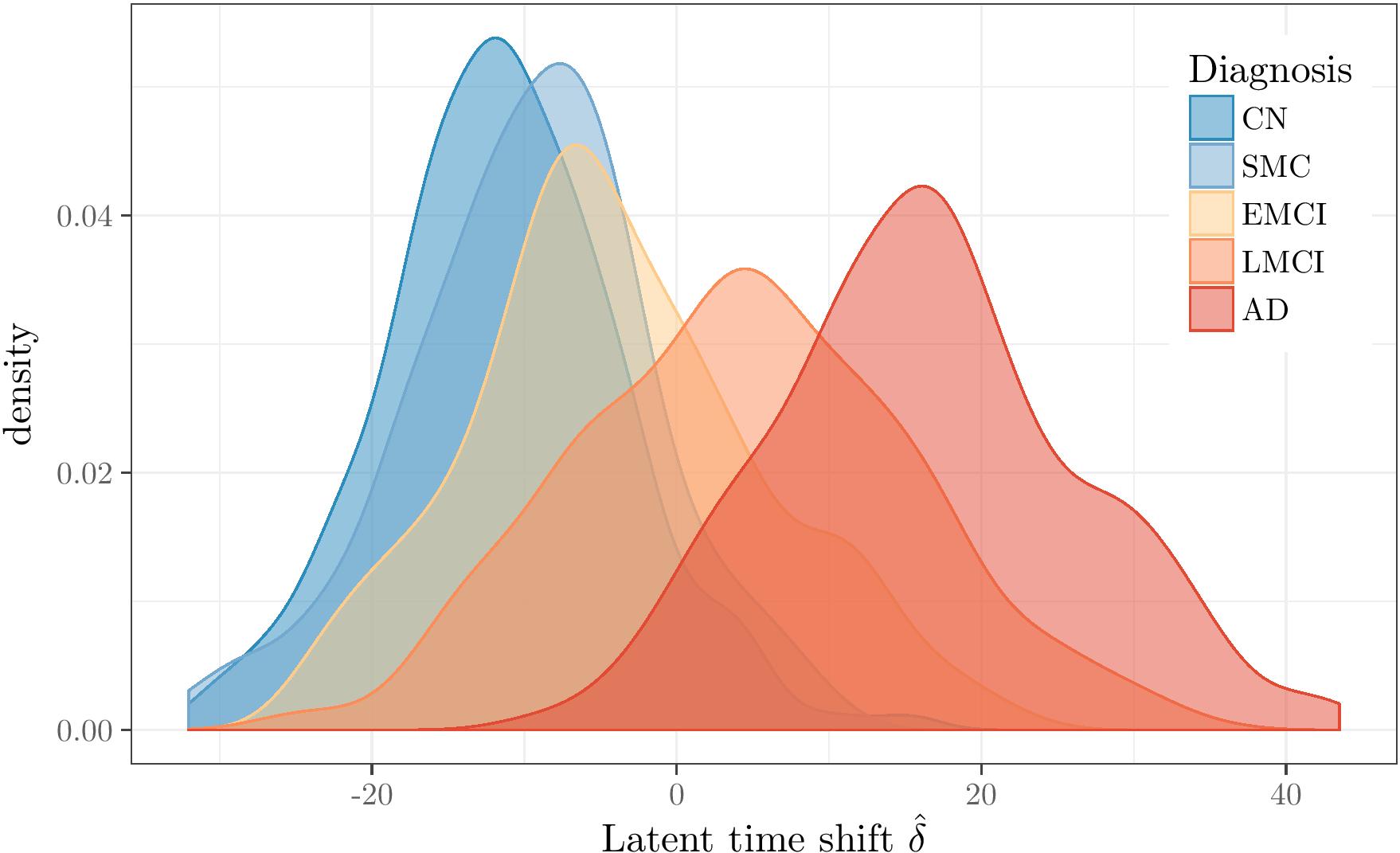}
	\caption[]{Distribution of the subject-specific latent time shifts. The estimated latent time shifts are colored by baseline diagnostic group, a variable not included in the model. This plot suggests that the time shifts are well aligned and consistent with diagnosis. The density plot also demonstrates that there is much overlap of the diagnostic criteria with respect latent time.\\
		\footnotesize{Abbreviations: CN, cognitively normal; SMC, subjective memory concern; EMCI, early mild cognitive impairment; LMCI, late mild cognitive impairment; AD, probable Alzheimer's Disease with mild to moderate dementia.} \label{latent_times}}
\end{figure}

Figure \ref{spaghetti} shows the subject-level observations (top) and predictions (bottom) according to age. It is clear from the observations, that age explains only a small proportion of the variance in these outcomes. The bottom panel shows that the predictions provide a reasonable smooth of the observations and that latent time provides a reasonable ordering of individuals according to disease severity. The posterior mean (95\% credible interval) for the latent time parameter was 14.2 (12.7 to 16.1) years (Table \ref{results_adni}). Figure \ref{latent_times} shows a density plot for the posterior mean of the subject-specific latent time by diagnosis at first ADNI visit. The latent time estimates are temporally sorting individuals in a manner that is consistent with physician diagnosis. Latent time parameters provide a continuous alternative to diagnosis which is objectively derived from a comprehensive model of longitudinal measures of disease.

\begin{figure}[H]
	\centering
	\subfigure[{Correlation between random intercepts.}]{\includegraphics[scale=0.4]{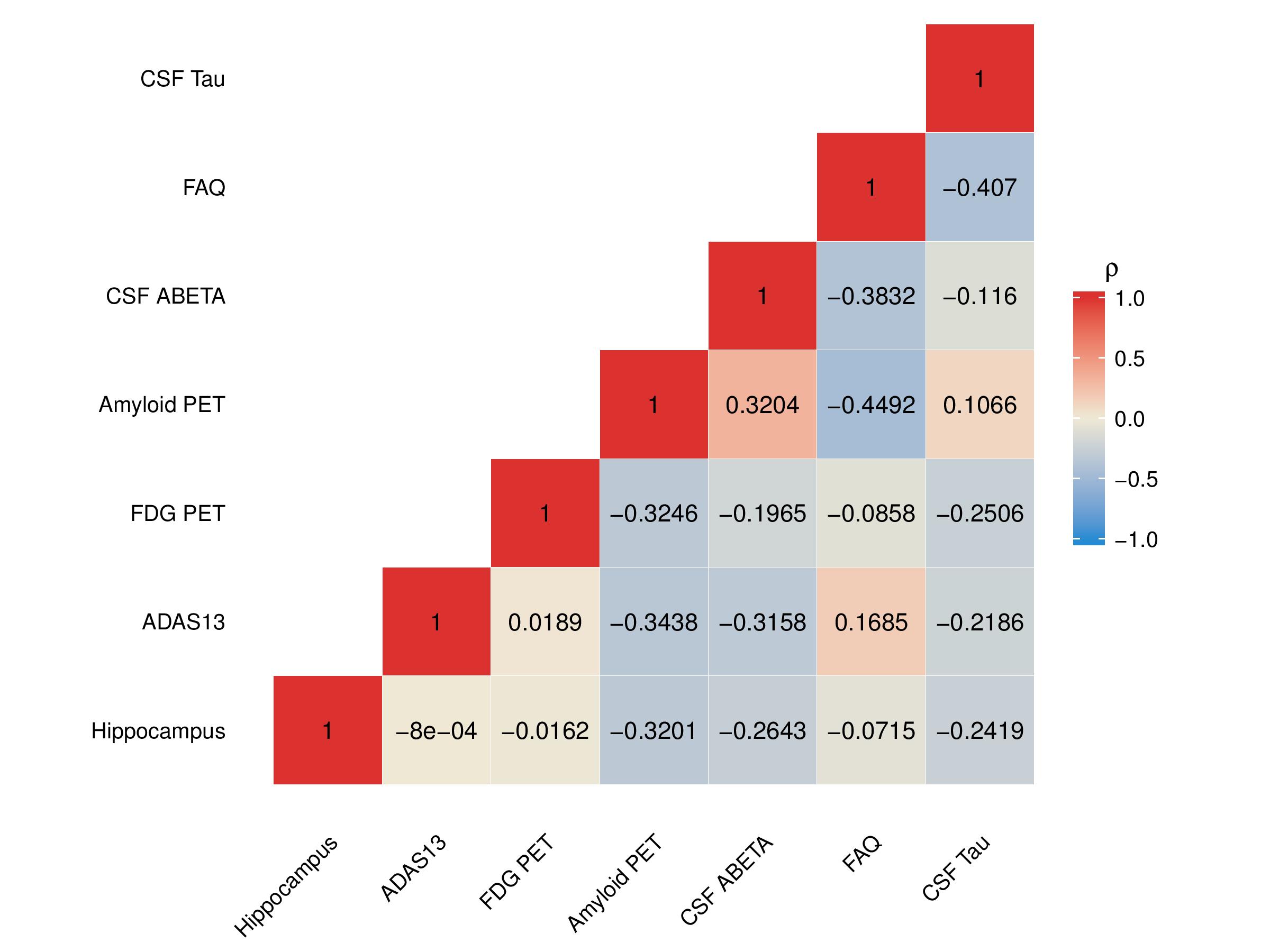}}
	\subfigure[{Correlation between random slopes.}]{\includegraphics[scale=0.4]{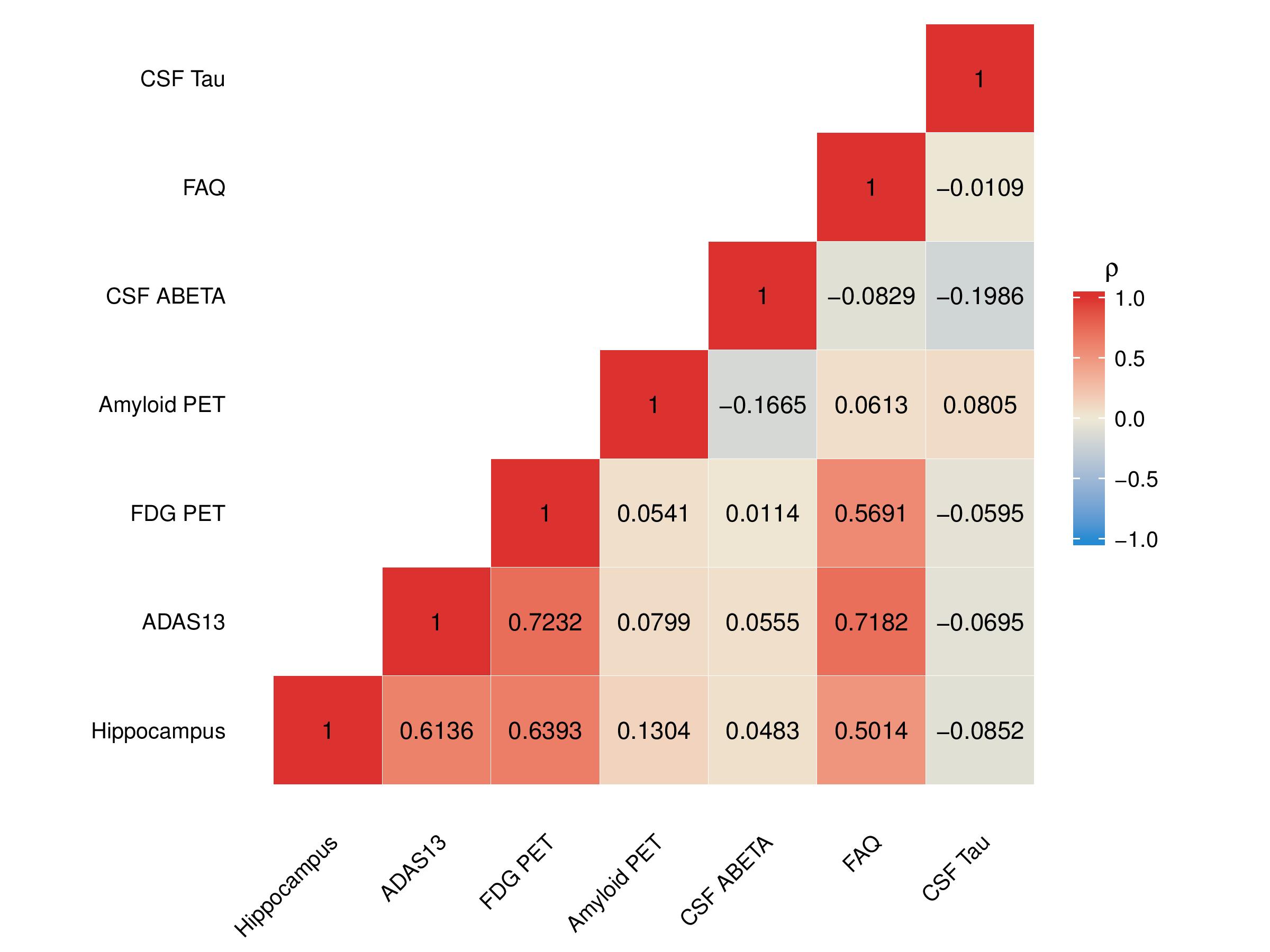}}
	\caption[]{Posterior mean correlations among random intercepts (a) and slopes (b).\\
		\footnotesize{Abbreviations: ADAS13, Alzheimer's Disease Assessment Scale (13 Item version); 
			FDG, fluorodeoxyglucose; PET, positron emission tomography; CSF, cerebrospinal fluid; FAQ, Functional Activities Questionnaire; ADNI, Alzheimer's Disease Neuroimaging Initiative.} \label{correlation_heatmap}}
\end{figure}

Figure \ref{correlation_heatmap} shows the posterior mean of correlation parameters for random intercepts and slopes. Not surprisingly, we see strong correlations between change in cognitive tests (ADAS13) and the function (FAQ). However we also, see strong correlations between measures of symptomatic change (cognition and function) and biomarkers (hippocampal atrophy and glucose metabolism [FDG PET]). The two amyloid measures, CSF A$\beta$ and amyloid PET, show a moderate positive correlation for random intercepts and weakly negative correlation of change.

\begin{figure}[H]
	\centering
	\includegraphics[scale=0.7]{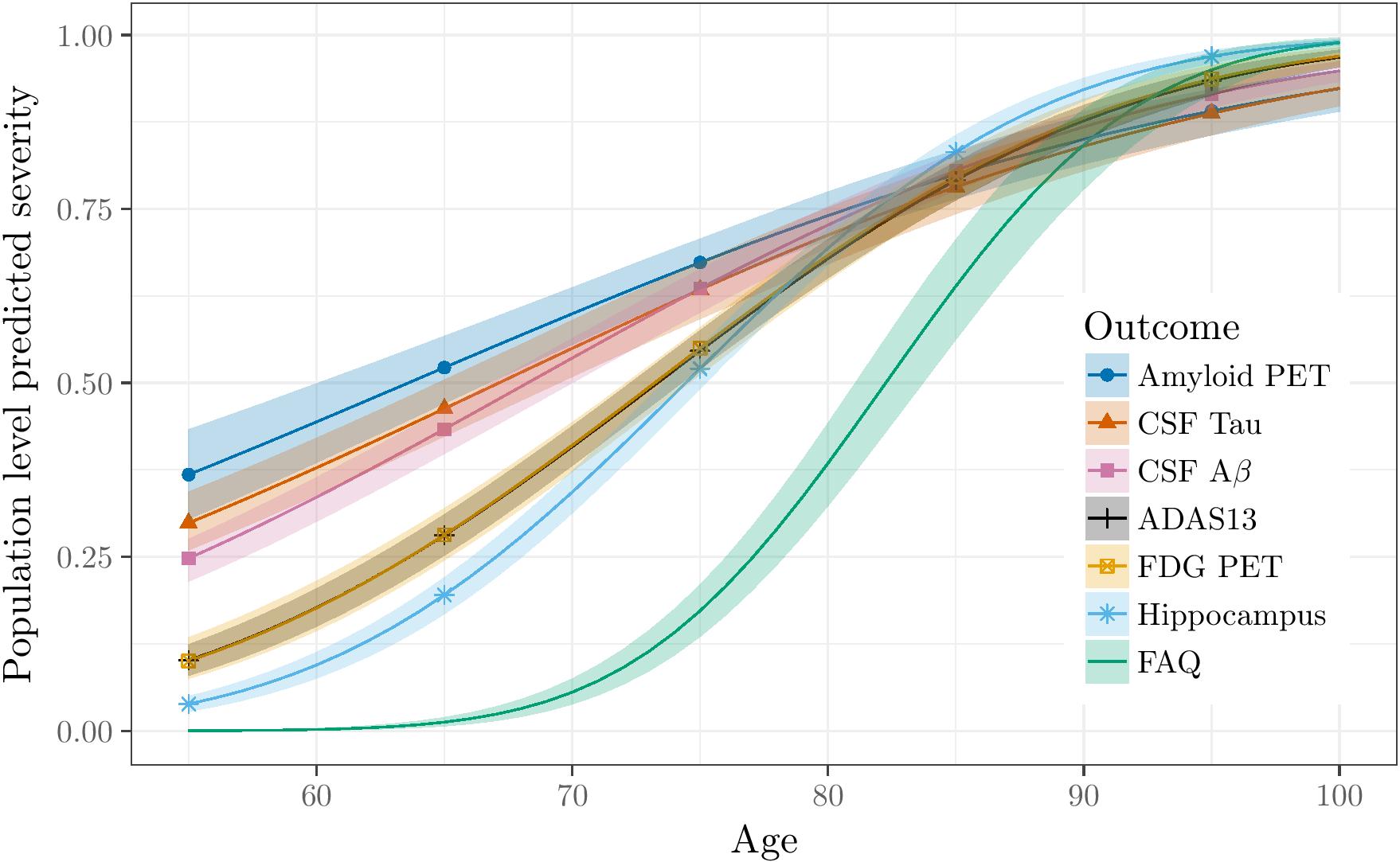}\\
	\vspace{10pt}
	\includegraphics[scale=0.7]{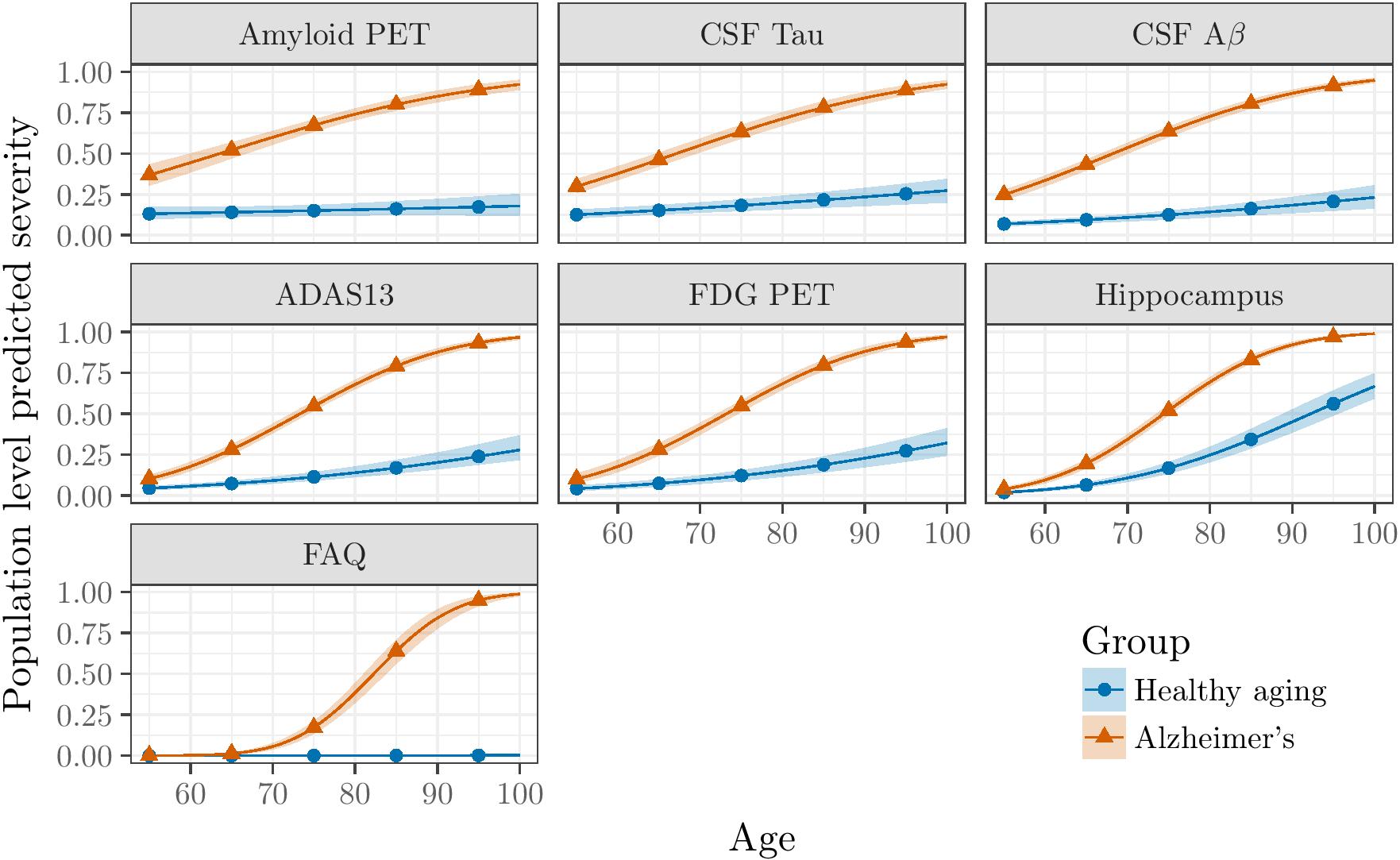}
	\caption[]{Modeled population-level severity. The top panel shows the modeled average trajectories. The depicted evolution is for female APOEe4 carriers with the ADNI mean education. Age is calibrated so that the ADAS13 trajectory attains the ADNI mean ADAS13 at the ADNI mean age at baseline. The bottom panel shows the same trajectories for progressive Alzheimer’s (red triangles) with contrasting trajectories for healthy aging (blue dots). For the latter, the effect of latent time is forced to be zero to isolate the effect of age.\\
		\footnotesize{Abbreviations: ADAS13, Alzheimer's Disease Assessment Scale (13 Item version); 
			FDG, fluorodeoxyglucose; PET, positron emission tomography; CSF, cerebrospinal fluid; FAQ, Functional Activities Questionnaire; ADNI, Alzheimer's Disease Neuroimaging Initiative.} \label{jack_curves}}
\end{figure}

Figure \ref{jack_curves} displays the population-level predicted trajectories. The depicted curves are for female APOEe4 carriers with the ADNI mean education. Recall that age and latent time contribute to the model independently. For the sake of these predictions, age is calibrated so that the ADAS13 trajectory attains the ADNI mean ADAS13 at the ADNI mean age at the first visit. The bottom panel shows the same trajectories for progressive Alzheimer's (red triangles) with contrasting trajectories for healthy aging (blue dots). To obtain estimates for healthy aging, the effect of latent time is forced to be zero to isolate the effect of age.

\begin{table}[H]
	\small\sf\centering
	\renewcommand{\arraystretch}{0.8}
	\caption{Results of posterior estimate of parameters for the proposed LTJMM fit to seven outcomes from the Alzheimer's Disease Neuroimaging Initiative (ADNI). \label{results_adni}} 
	\vspace{8pt}
	{\tabcolsep = 4pt
		\begin{tabular}{crc|crc}
			\hline
			Parameter & \multicolumn{1}{c}{Posterior} & 95\% & Parameter & \multicolumn{1}{c}{Posterior} & 95\% \\
			& \multicolumn{1}{c}{Mean} & Credible Interval &  & \multicolumn{1}{c}{Mean} & Credible Interval\\ \hline
			\multicolumn{3}{c|}{\textbf{Hippocampal volume}} & \multicolumn{3}{c}{\textbf{CSF Amyloid Beta 1-42}} \\
			Intercept & -4.102 & $\left(-4.516,-3.688\right)$       & Intercept & -1.454 & $\left(-2.061,-0.872\right)$ \\
			Age       & 0.056 & $\left(0.052,0.06\right)$           & Age       & 0.016 & $\left(0.011,0.022\right)$ \\
			APOE$\varepsilon$4 & 0.317 & $\left(0.258,0.394\right)$ & APOE$\varepsilon$4 & 0.802 & $\left(0.704,0.888\right)$ \\
			Sex       & -0.288 & $\left(-0.364,-0.215\right)$       & Sex & -0.028 & $\left(-0.116,0.049\right)$ \\
			Education & -0.003 & $\left(-0.015,0.008\right)$        & Education & -0.012 & $\left(-0.03,0.005\right)$ \\
			Latent time & 0.035 & $\left(0.03, 0.04\right)$         & Latent time & 0.035 & $\left(0.03,0.039\right)$ \\
			$\sigma_1$ & 0.137 & $\left(0.134, 0.141\right)$        & $\sigma_{5}$ & 0.021 & $\left(0.019,0.022\right)$ \\[3pt]
			\multicolumn{3}{c|}{\textbf{Alzheimer's Disease Assessment Scale}} & \multicolumn{3}{c}{\textbf{Functional Activities Questionnaire}} \\
			Intercept & -1.089 & $\left(-1.57, -0.646\right)$       & Intercept & -3.012 & $\left(-3.934,-2.012\right)$\\
			Age       & 0.025 & $\left(0.02, 0.03\right)$           & Age       & 0.035 & $\left(0.025,0.045\right)$\\
			APOE$\varepsilon$4 & 0.434 & $\left(0.355, 0.514\right)$ & APOE$\varepsilon$4 & 0.861 & $\left(0.678,1.021\right)$\\
			Sex       & -0.246 & $\left(-0.314, -0.168\right)$      & Sex & -0.335 & $\left(-0.471,-0.196\right)$\\
			Education & -0.051 & $\left(-0.064,-0.038\right)$       & Education & -0.061 & $\left(-0.089,-0.037\right)$\\
			Latent time & 0.045 & $\left(0.039, 0.05\right)$        & Latent time & 0.096 & $\left(0.085,0.107\right)$\\
			$\sigma_2$ & 0.327 & $\left(0.32, 0.333\right)$         & $\sigma_6$ & 0.842 & $\left(0.826,0.858\right)$\\[3pt]
			\multicolumn{3}{c|}{\textbf{FDG PET}} & \multicolumn{3}{c}{\textbf{CSF Tau}} \\
			Intercept & -1.820 & $\left(-2.348, -1.3\right)$         & Intercept & -0.766 & $\left(-1.265,-0.202\right)$\\
			Age       & 0.028 & $\left(0.023, 0.034\right)$         & Age       & 0.012 & $\left(0.007,0.017\right)$ \\
			APOE$\varepsilon$4 & 0.453 & $\left(0.363, 0.54\right)$ & APOE$\varepsilon$4 & 0.621 & $\left(0.505,0.72\right)$\\
			Sex       & -0.181 & $\left(-0.259, -0.103\right)$      & Sex & 0.073 & $\left(-0.036,0.19\right)$\\
			Education & -0.026 & $\left(-0.041,-0.01\right)$        & Education & -0.029 & $\left(-0.048,-0.007\right)$\\
			Latent time & 0.042 & $\left(0.037, 0.048\right)$       & Latent time & 0.032 & $\left(0.028,0.036\right)$ \\
			$\sigma_3$ & 0.294 & $\left(0.283, 0.304\right)$        & $\sigma_{7}$ & 0.016 & $\left(0.014,0.017\right)$ \\[3pt]
			\multicolumn{3}{c|}{\textbf{Amyloid PET}} & \multicolumn{3}{c}{\textbf{Standard deviation of latent time}}  \\
			Intercept & -0.685 & $\left(-1.328, -0.013\right)$      & $\sigma_{\delta}$ & 14.262 & $\left(12.684,16.105\right)$\\
			Age       & 0.004 & $\left(-0.002, 0.011\right)$ & & & \\
			APOE$\varepsilon$4 & 0.785 & $\left(0.669, 0.912\right)$ & & & \\
			Sex       & 0.159 & $\left(0.067, 0.247\right)$ & & & \\
			Education & -0.007 & $\left(-0.029,0.014\right)$ & & & \\
			Latent time & 0.035 & $\left(0.03, 0.04\right)$ & & & \\
			$\sigma_4$ & 0.289 & $\left(0.273, 0.307\right)$ & & & \\
			\hline
		\end{tabular}}	
\end{table}

\begin{table}[H]
	\small\sf\centering
	\renewcommand{\arraystretch}{0.8}
	\caption{Results of model selection for models fit to seven outcomes from the Alzheimer's Disease Neuroimaging Initiative (ADNI). \label{results_adni_b}} 
	\vspace{8pt}
		{\tabcolsep = 4pt
			\begin{tabular}{@{}l|ccc@{}}
				\hline
				Model & WAIC & LOOIC & DIC \\ \hline 
				$\textrm{LTJMM}_1$ & 10246.31 & 14156.73 & 11070.65 \\  
				$\textrm{LTJMM}_2$ & 9684.75 & 13482.41 & 10528.48 \\ 
				$\textrm{MM}$ & 10023.88 & 13938.55 & 11010.95 \\  
				$\textrm{JMM}$ & 9308.89 & 12385.48 & 10170.97 \\ 
				$\textrm{MM (diagnosis)}$ & 9792.05 & 13778.70 & 10761.32 \\
				$\textrm{JMM (diagnosis)}$ & 9420.06 & 13082.31 & 10286.82 \\
				\hline
			\end{tabular}
			\begin{tablenotes}
			\footnotesize
			\item Note: Model comparison results for LTJMMs with univariate Gaussian random effects ($\textrm{LTJMM}_1$) and with multivariate Gaussian random effects ($\textrm{LTJMM}_2$); conventional linear mixed-effects model (MM); joint mixed effects model (JMM) with multivariate Gaussian random effects; and MM (diagnosis) and JMM (diagnosis) with categorical baseline diagnosis and its interaction with time as additional covariates.
			\end{tablenotes}}
\end{table}

\section{Discussion}
\label{sec4}

We explored sampling the random effects from both univariate and multivariate Gaussian distributions, and found the multivariate is always preferred when it is the true model, and performs similarly well when the univariate is the true model. The model with multivariate random effects has the advantage of providing additional parameters for the correlation among outcomes (random intercepts) and among change in outcomes (random slopes). However, fitting the model with univariate random effects was often computationally faster, so it may have advantages in practice. We plan to explore extension of the model which include monotone smooth functions of latent time, and other random effects options. Furthermore, the variability of latent time shifts was assumed to be homogeneous (same for all individuals) in this work. We are currently working on an extension of the model to allow heterogeneous variances by explicitly modeling subject-specific latent time variance in terms of observed covariates.

Interestingly, the model selection criteria preferred LTJMM over MM when the MM did not include fixed effects for categorical baseline diagnosis. However, when the MM had the advantage of the additional covariates for baseline diagnosis, the MM was preferred over LTJMM. This analysis suggests LTJMM is a good alternative to MM when baseline diagnosis is not known or when novel insights about long-term progression across all the diagnostic categories are desired. These long-term insights are unattainable with MM. LTJMM also provides subject-specific latent disease-time estimates that could serve as a continuous objective alternative to categorical diagnosis.

The LTJMM provides a parsimonious extension of the existing family of joint generalized linear mixed effects models with a subject-specific latent time parameter that accommodates the temporal heterogeneity common to multicohort longitudinal data in late-life neurodegenerative diseases. The model provides key insights and inference regarding the evolution of disease markers over a period of time that is longer than the period of observation. The framework allows consideration of the independent effects of healthy aging and disease progression. 

These novel features can be leveraged to improve subject-level prediction and better understand long-term disease dynamics. In particular we plan to leverage the model to help improve clinical trial design. For example, given an estimate of the short-term effect of a treatment on a biomarker, the model can be interrogated to estimate the downstream effects on cognition and function. The model can also be used to help identify populations expected to experience the maximum benefit from a given intervention. The LTJMM provides a flexible framework to begin to explore these and many other applications.

\section{Supplementary material}
\label{sec6}

The \textsf{Stan} code model specifications and related code and functions for simulation, estimation and prediction are available as an \textsf{R} package from \url{https://bitbucket.com/mdonohue/ltjmm}. Interactive convergence plots of the model fit to Alzheimer's data is available from \url{https://shiny.atrihub.org/public/adni_ltjmm/}.

\section*{Acknowledgment}

We are grateful to the ADNI study volunteers and their families. This work was supported by Biomarkers Across Neurodegenerative Disease (BAND-14-338179) grant from the Alzheimer's Association, Michael J. Fox Foundation, and Weston Brain Institute; and National Institute on Aging grant R01-AG049750. Data collection and sharing for this project was funded by the ADNI (National Institutes of Health Grant U01 AG024904) and DOD ADNI (Department of Defense award number W81XWH-12-2-0012). ADNI is funded by the National Institute on Aging, the National Institute of Biomedical Imaging and Bioengineering, and through generous contributions from the following: AbbVie, Alzheimer's Association; Alzheimer's Drug Discovery Foundation; Araclon Biotech; BioClinica, Inc.; Biogen; Bristol-Myers Squibb Company; CereSpir, Inc.; Eisai Inc.; Elan Pharmaceuticals, Inc.; Eli Lilly and Company; EuroImmun; F. Hoffmann-La Roche Ltd and its affiliated company Genentech, Inc.; Fujirebio; GE Healthcare; IXICO Ltd.; Janssen Alzheimer Immunotherapy Research \& Development, LLC.; Johnson \& Johnson Pharmaceutical Research \& Development LLC.; Lumosity; Lundbeck; Merck \& Co., Inc.; Meso Scale Diagnostics, LLC.; NeuroRx Research; Neurotrack Technologies; Novartis Pharmaceuticals Corporation; Pfizer Inc.; Piramal Imaging; Servier; Takeda Pharmaceutical Company; and Transition Therapeutics. The Canadian Institutes of Health Research is providing funds to support ADNI clinical sites in Canada. Private sector contributions are facilitated by the Foundation for the National Institutes of Health (www.fnih.org). The grantee organization is the Northern California Institute for Research and Education, and the study is coordinated by the Alzheimer's Therapeutic Research Institute at the University of Southern California. ADNI data are disseminated by the Laboratory for Neuro Imaging at the University of Southern California.\\

\bibliographystyle{elsarticle-num}
\bibliography{refs} 

\newpage
\appendix
\section{Proof of the Identifiability}

Suppose that $p$ response variables are measured for $n$ individuals at different follow-up times.  The measured outcomes for individual $i$ at time $j$ are denoted as $\by_{ij}=\left(y_{ij1},\cdots, y_{ijp}\right)^{\prime}$.   The latent time joint mixed effects model is given by
\begin{equation}
	\label{LTJMM}
	y_{ijk} = \bx^\prime_{ijk}\left(t_{ijk}\right) \bbeta_k+ \gamma_k \left(t_{ijk} + \delta_i\right) + \alpha_{0ik} + \alpha_{1ik} t_{ijk} + \varepsilon_{ijk},
\end{equation}
where $i=1,\ldots,n$, $k=1,\ldots,p$ and $j=1,\ldots,q_{ik}$.

The model (\ref{LTJMM}) can be re-expressed as the form of linear mixed effects model
\begin{equation}
	\label{lme}
	y_{ijk} = \bx^\prime_{ijk}\left(t_{ijk}\right) \bbeta_k + \gamma_k t_{ijk} + \tilde{\alpha}_{0ik} + \alpha_{1ik}t_{ijk} + \varepsilon_{ijk},
\end{equation}
where $\tilde{\alpha}_{0ik} = \gamma_k \delta_i + \alpha_{0ik}$. Identifiability, up to $\tilde{\alpha}_{0ik}$ follows from the identifiability of standard linear mixed effects models (see Wang (2013)\footnote{Wang, W. (2013). Identifiability of linear mixed effects models. \emph{Electronic Journal of Statistics}, 7, 244-263.} for a discussion) provided $(\bx^\prime, \bt)$ has full column rank, where $\bt$ is the vector formed by $t_{ijk}$; and $\bt$ is linearly independent from the vector of 1's. 

By placing the sum-to-zero constraints on $\alpha_{0ik}$, i.e., $\sum_{k=1}^p\alpha_{0ik}=0$, it remains to show the system below:
\begin{align}
	\label{eq3}
	\tilde{\alpha}_{0ik} &= \gamma_k \delta_i + \alpha_{0ik}\\
	\label{eqc1}
	0 &= \sum_{k=1}^p \alpha_{0ik}
\end{align}
has a unique solution, where $\gamma_k$ and $\tilde{\alpha}_{0ik}$ are known, since they are identifiable in (\ref{lme}); $\delta_i$ and $\alpha_{0ik}$ are unknown. There are $n \times p$ unknown subject and outcome specific parameters $\alpha_{0ik}$, and $n$ subject-specific latent time parameters $\delta_i$, for a total of $np+n$ unknown parameters. There are $np+n$ constraints (or equations). The system (\ref{eq3} \& \ref{eqc1}) with the same number of equations and unknown parameters has a unique solution. Therefore, $\delta_i$ and $\alpha_{0ik}$ are uniquely determined. Thus the model is identifiable.

\end{document}